# Data acquisition system for a 146-channel counter of protons in particle therapy

S. Giordanengo[1], M. Abujami[1,2], C. Galeone[2], S. Garbolino[1], O. A. Martì Villarreal[1,2], F. Mas Milian[1,2,3], G. Mazza[1], M. Mignone[1], A. Vignati[1,2], R. Wheadon[1], R. Cirio[1,2], V. Monaco[1,2], R. Sacchi[1,2]

1. National Institute for Nuclear Physics INFN, Turin division, Turin, Italy
2. Università degli Studi di Torino, Physics Department, Turin, Italy
3. Universidade Estadual de Santa Cruz, Ilheus, Brazil

*Abstract*— A prototype of proton counter was developed by the University and the National Institute for Nuclear Physics of Torino to be used as online fluence beam monitor in particle therapy. The single particle identification approach aims at increasing the sensitivity and readout speed with respect to the state-of-the-art gas ionization chambers. The sensitive area is 2,7×2,7 cm$^2$ to cover the clinical beam cross section characterized by a full width at half maximum of about 1 cm at the isocenter. The sensor is a thin Low Gain Avalanche Diode segmented in 146 strips with 180 um pitch and with 50 μm active thickness, designed and produced by the Fondazione Bruno Kessler (Trento, Italy). The frontend readout to identify the single proton signal provided by each strip is based on a 24-channel custom ASICs, named ABACUS, optimized to discriminate the signal pulses in a wide charge range (3-150 fC) with a maximum dead-time of 10 ns. With these specifications, at the maximum fluence rate of $10^8$ p/(cm$^2 \cdot$s) in the clinical energy range (60-230 MeV) and considering the silicon strips described above, a maximum pileup counting inefficiency < 1% is achieved. A frontend board housing 6 ABACUS chips to readout the 146 strips was developed, the digital outputs being sent to 3 FPGAs (Kintex7) for the counting. A LabVIEW program implements the interface with the FPGAs, displays online the counting rate from each strip and stores the data for offline analysis.

The proton counter data acquisition system will be presented and the preliminary performances in terms of baseline uniformity, noise, linearity with charge, efficiency with pulse rate and data throughput will be reported.

## I. INTRODUCTION

Among available cancer treatment techniques, particle therapy has great potentials in reducing toxicities and improving quality of life [1]. However, the clinical implementation of advanced delivery techniques (e.g. fast rescanning modalities) is prevented by the limits of the currently used beam monitors (gas-filled ionization chambers - IC) based on real-time integration of the charge produced in the gas by the delivered particles. The existing ICs suffer from several important limitations [2]. Because of the small ionization charge produced in gases, their sensitivity is limited to the order of thousands of particles. Additionally, slow charge collection times (hundreds of microseconds) and charge recombination effects preclude their use on fast beam delivery strategies, which represent the trend for future radiotherapy approaches, to improve the accuracy, to shorten the treatments and to increase the patient throughput.

To overcome these limits, the University and National Institute of Nuclear Physics developed a pioneer device based on innovative silicon detectors integrated with custom Very Large Scale Integration (VLSI) chips and a dedicated data acquisition architecture to prove the feasibility of counting the number of delivered particles. The silicon sensor, the custom frontend readout and the whole data acquisition system will be presented together with the preliminary performances evaluated with tests performed both in laboratory and with clinical beam.

## II. THE SILICON SENSOR

Prototypes of silicon sensors based on Low Gain Avalanche Diode (LGAD) technology [3] have been designed and produced by the Fondazione Bruno Kessler (FBK, Trento, Italy) in the framework of the MoVeIT INFN project [4]. Thin planar silicon sensors have been chosen because they ensure charge uniformity collection, and the reduced thickness (tens of micrometers) guarantees short signals duration and low material budget after thinning. Additionally, a low internal gain leads to signal enhancement with similar noise level of a traditional silicon sensor of the same geometry. The optimized sensors for a proton counter feature 50 μm of active thickness, 146 strips 2.7 cm long with 180 μm pitch to achieve a 2.7 x 2.7 cm$^2$ sensitive area. The segmentation allows to reduce the expected particle rate per channel and the sensor capacitance while the sensitive area was defined to cover the clinical proton beam cross section characterized by a full width at half maximum of about 1 cm at the isocenter.

## III. FRONTEND READOUT AND DATA ACQUISITION

The frontend readout developed to identify the single proton signal provided by each strip at a frequency up to 100 MHz, is based on 24-channel custom ASIC, designed to discriminate the expected signal pulses in a wide charge range (3-150 fC) with a maximum dead-time of 10 ns [5]. These specifications come from the clinical fluence rate ($10^6$-$10^{10}$ protons/cm$^2 \cdot$s), clinical energy range (60-230 MeV) and considering the silicon strips described above. Two flavours of ASIC have been developed in a commercial CMOS 0.11 μm technology: one is a high bandwidth Trans-Impedance Amplifier based on a differential amplifier with resistive load, gain boost current sources, and a resistive feedback. The second architecture, named ABACUS, is a fast Charge Sensitive Amplifier described in [5]. A frontend board, named ESA_ABACUS, was developed to house six ABACUS chips and readout the 146 strips as shown in Fig.2. Two strips are connected to two analog outputs on the board for debug and the others 144 strips are wired-bonded to the six ABACUS input channels.

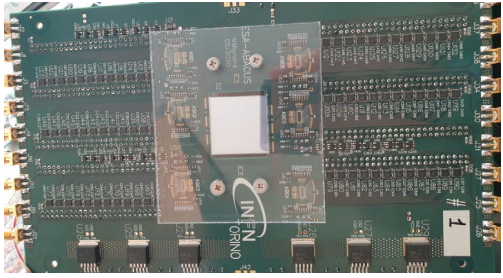

*Fig. 1 Picture of the ESA_ABACUS frontend board*

The discriminator output for both architectures drives differential CML signals to maintain high signal rate; on the board these are converted into LVDS and sent to FPGAs for the pulses counting. Three Xilinx Kintex7 (KC705 Evaluation Kit) FPGAs with main clock of 100 MHz and deserializer working at 1 GHz are used to readout 144 channels. A Gigabit ethernet interface connects each FPGA with a computer where a LabVIEW program controls and integrates all the FPGAs input-outputs, displays online the counting rate from each strip and stores the useful data like timestamp, raw integral counts and thresholds in files for later analysis. Three onboard DACs allow to set a global threshold (one for each chip) and local DACs have been implemented per channel in the chip to fine tuning of the comparator threshold needed to obtain a uniform response of the strips. The setting of DACs is performed through the FPGAs outputs driven by the LabVIEW software.

Non linearity effects due to signal overlapping are mitigated with robust algorithms, based on logical combinations of signals from neighboring strips, implemented in the FPGAs.

## IV. RESULTS

Tests of LGAD silicon sensors performed on synchrotron and cyclotron beams of therapeutic centers, using a pin-hole ionization chamber for the independent measurement of the particle flux, show the possibility to keep the counting error < 1 % up to a beam fluxes of $5 \cdot 10^8$ protons/(cm$^2 \cdot$s).

An extensive characterization work of the ESA_ABACUS frontend board was performed with the ASICs coupled to the sensor evaluating for each channel and chip the baseline uniformity, noise, linearity with charge, and efficiency with pulses rate. Tests were done both in laboratory with external pulse generator used to inject different charges through dedicated test pulser and with therapeutic proton beams. Fig. 2 shows as example the linearity of the pulse amplitude as a function of the injected charge for 7 channels of chip 6, which allows to determine the pedestal and gain of each channel.

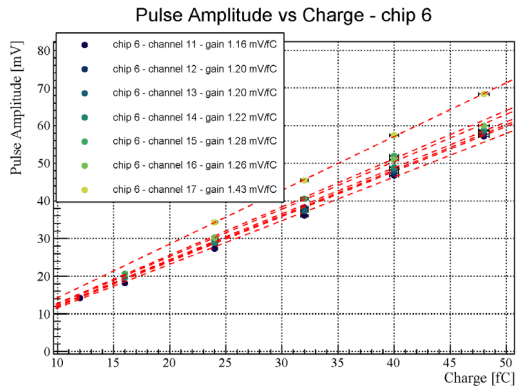

*Fig. 2 Linearity and gain of 7 channels of chip 6 vs charge.*

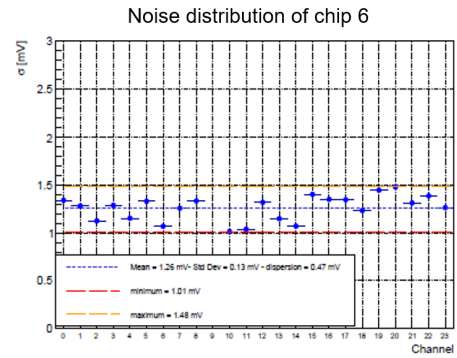

*Fig. 3 Noise distribution for 24 channels of chip 6 connected to 24 strips of the 146-strip sensor with a bias voltage of 200 V.*

The noise distribution was determined from the data and is shown in Fig. 3. It is found to be fairly uniform between channels and ranges between 1 and 1.5 mV ensuring the Signal to Noise Ratio in excess of 10 at 10 fC as you can be seen comparing Fig. 2 and Fig. 3.

The detector data acquisition performance will be also presented in terms of maximum rate of measurements and counting efficiency.

## V. CONCLUSIONS

A prototype of proton counter to be used as online fluence beam monitor was developed to overcome the limits of the state-of-the-art gas ionization chambers. The preliminary tests of the final detector based on a silicon detector segmented in 146 strips and dedicated DAQ system prove the feasibility of a direct measurement of the particle rate during a treatment with protons. Towards the use of this technology as beam monitor in the clinical practice, further studies will be done to improve radiation resistance, use finer segmentation and increase the detector sensitive area.

## ACKNOWLEDGMENT

This work has been performed within the framework of the MoVeIT project, founded by the INFN-CSN5, and has been supported by MIUR Dipartimenti di Eccellenza (ex L.232/2016, art.1, cc. 314, 337). FMM was supported by the FAPESB fellowship. Part of this work has been also supported by the European Union's Horizon 2020 Research and Innovation funding program (Grant Agreement no. 669529 - ERC UFSD669529).